# Stress dependent thermal pressurization of a fluid-saturated rock

Siavash Ghabezloo[1] and Jean Sulem

Université Paris-Est, UR Navier-CERMES, Ecole Nationale des Ponts et Chaussées

## Abstract

Temperature increase in saturated porous materials under undrained conditions leads to thermal pressurization of the pore fluid due to the discrepancy between the thermal expansion coefficients of the pore fluid and of the solid matrix. This increase in the pore fluid pressure induces a reduction of the effective mean stress and can lead to shear failure or hydraulic fracturing. The equations governing the phenomenon of thermal pressurization are presented and this phenomenon is studied experimentally for a saturated granular rock in an undrained heating test under constant isotropic stress. Careful analysis of the effect of mechanical and thermal deformation of the drainage and pressure measurement system is performed and a correction of the measured pore pressure is introduced. The test results are modelled using a non-linear thermo-poro-elastic constitutive model of the granular rock with emphasis on the stress-dependent character of the rock compressibility. The effects of stress and temperature on thermal pressurization observed in the tests are correctly reproduced by the model.

[1] Corresponding author, CERMES, Ecole Nationale des Ponts et Chaussées, 6-8 avenue Blaise Pascal, Cité Descartes, 77455 Champs-sur-Marne, Marne la Vallée cedex 2, France, Email : ghabezlo@cermes.enpc.fr

# 1 Introduction

Temperature increase in saturated porous materials leads to thermal pressurization of the pore fluid due to the discrepancy between the thermal expansion coefficients of the pore fluid and of the solid phase. This increase in the pore fluid pressure induces a reduction of the effective mean stress, and can lead to shear failure or hydraulic fracturing. This phenomenon is important in petroleum engineering where the reservoir rock or the well cement lining undergoes sudden temperature changes as well as in environmental engineering for radioactive waste disposal in deep clay geological formations. It is also important in geophysics in the studies of rapid fault slip events when shear heating tends to increase the pore pressure and to decrease the effective compressive stress and the shearing resistance of the fault material (Rempel and Rice 2006, Sulem et al. 2007). Important theoretical advances have been proposed in the study of thermal weakening of fault during coseismic slip and one can find an extensive literature review on the subject in the comprehensive paper of Rice (2006). In this paper, Rice emphasises the need of laboratory data to constrain theoretical modelling of these mechanisms. In particular thermal pressurization of rocks during seismic slip is highly influenced by damage and inelastic deformation inside the fault zones. The presence of clay material in fault zones also affects thermal pressurization as possible collapse of the clay under thermal loading may activate fluid pressurization (Sulem et al. 2007).

The values of the undrained thermal pressurization coefficient, defined as the pore pressure increase due to a unit temperature increase in undrained condition, is thus largely dependent upon the nature of the material, the state of stress, the range of temperature change, the induced damage. In the literature we can find values that differ from two orders of magnitude: In Campanella and Mitchell (1968) different values are found from 0.01 MPa/°C for clay to 0.05 MPa/°C for sandstone. Palciauskas and Domenica (1982) estimate a value of 0.59MPa/°C for Kayenta sandstone. On the basis of Sultan (1997) experimental data on Boom clay, Vardoulakis (2002) estimates this coefficient as 0.06 MPa/°C. For a clayey fault gouge extracted at a depth of 760m in Aigion fault in the Gulf of Corinth (Greece), the value obtained by Sulem et al. (2004) is 0.1 MPa/°C and for intact rock at great depth, the value given by Lachenbruch (1980) is 1.5 MPa/°C. For a mature fault at 7km depth at normal stress of 196 MPa, ambient pore pressure of 70 MPa, and ambient temperature of 210°C, Rice (2006) estimates this coefficient as 0.92MPa/°C in case of intact fault walls and 0.31MPa/°C in case of damage fault wall.

The large variability of the thermal pressurization coefficient highlights the necessity of laboratory studies. The aim of this paper is to study the phenomenon of thermal pressurization, theoretically and experimentally for a saturated granular rock. The stress dependency of the compression modulus of granular rocks has important consequences on the thermal response in undrained condition and this phenomenon is addressed here. In the treatment of the test results, the effect of mechanical and thermal deformation of the drainage system is carefully analysed and taken into account.

# 2 Thermo-poro-mechanical background

The equations governing the phenomenon of the thermo-mechanical pressurization of porous materials are presented here for an ideal porous material characterized by a fully connected pore space and by a microscopically homogeneous (i.e., composed of only one solid material) and isotropic solid phase. These equations can be derived using the approach of Bishop and Eldin (1950). In this approach, schematized on Figure (1), the problem of thermo-mechanical loading is broken up into three independent sub-problems. Assuming isotropic elasticity, the volumetric changes of the porous material element and its components, are written separately, as presented on table (1). In this table,

- $V$ is the total volume
- $n$ is the Lagrangian porosity,
- $\Delta\sigma$, $\Delta T$ and $\Delta u$ are the variations of the mean stress (positive in compression), the temperature and the pore pressure respectively,
- $c_f$ and $c_s$ are respectively the compressibility of pore fluid and of the solid phase,
- $c_d$ is the drained compressibility of the porous material,
- $\alpha_f$ and $\alpha_s$ are respectively the volumetric thermal expansion coefficient of pore fluid and solid phase.

The index $0$ denotes the reference state.

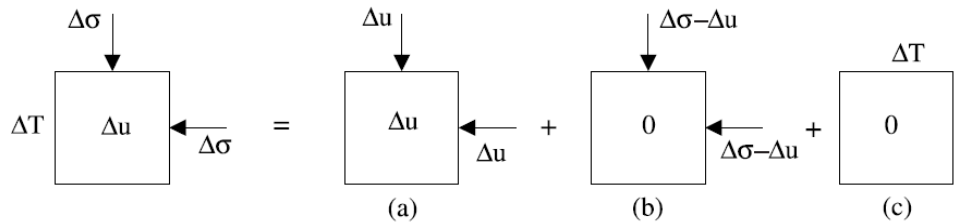

**Figure 1: Decomposition of the problem of a thermo-mechanical loading into three independent problems**

| Problem | Variation of volume of constituents | | Variation of volume of element |
|---|---|---|---|
| | Pore fluid | Solid phase | |
| (a) | $-n_0 V_0 c_f \Delta u$ | $-(1-n_0)V_0 c_s \Delta u$ | $-V_0 c_s \Delta u$ |
| (b) | - | $-V_0 c_s (\Delta\sigma - \Delta u)$ | $-V_0 c_d (\Delta\sigma - \Delta u)$ |
| (c) | $n_0 V_0 \alpha_f \Delta T$ | $(1-n_0) V_0 \alpha_s \Delta T$ | $V_0 \alpha_s \Delta T$ |

**Table 1: Variation of volume of the porous material and its constituents for the three independent problems presented in Figure (1)**

A clear distinction must be made between the Lagrangian and the Eulerian porosity (Rice and Cleary 1976, Coussy 2004). The Lagrangian porosity $n$ is defined as the pore volume per unit volume of porous material in the reference state (also called pore volume fraction), while the Eulerian porosity $\phi$ is defined as the pore volume per unit volume of porous material in actual state. Both quantities are identical in the reference state ($n_0 = \phi_0$) but differ in the actual (deformed) state. The following relations hold for the Lagrangian and Eulerian porosity change:

$$n - n_0 = \frac{V_n - V_{n0}}{V_0} \tag{1}$$

$$\phi - \phi_0 = \frac{V_n}{V} - \frac{V_{n0}}{V_0} \tag{2}$$

$$n - n_0 = \phi - \phi_0 + \phi\left(\frac{V}{V_0} - 1\right) \tag{3}$$

where $V_n = nV_0$ is the volume of the pore space in actual state and $V_{n0} = n_0 V_0$ is the volume of the pore space in reference state.

The problem (a) in the Figure (1) corresponds to a particular loading case, called 'unjacketed' loading case, where an isotropic stress and a pore pressure of equal magnitudes are simultaneously applied to a volume element. This loading is as if the sample is submerged, without a jacket, in a fluid under pressure and results in a uniform pressure distribution in the solid phase. In the case of an ideal porous material, the material would deform as if all the pores were filled with the solid component. The skeleton and the solid component experience a uniform volumetric strain with no change of the Eulerian porosity $\phi$ (Detournay and Cheng 1993).

$$\frac{\Delta V_s}{V_s} = \frac{\Delta V_n}{V_n} = \frac{\Delta V}{V} \qquad (4)$$

The problem (b) in the Figure (1) corresponds to Terzaghi effective stress loading. In this case the isotropic stress acting on the solid phase is $(\Delta\sigma - \Delta u)/(1-n_0)$ and the corresponding variation of volume is given by:

$$\Delta V_s = -(1-n_0)V_0 c_s \frac{\Delta\sigma - \Delta u}{1-n_0} = -V_0 c_s (\Delta\sigma - \Delta u) \qquad (5)$$

The problem (c) in the Figure (1) is a drained thermal loading under constant total stress. For an ideal porous material, the variation of volume of the porous element is characterized by the thermal expansion coefficient of the solid phase $\alpha_s$. In this case no change in Eulerian porosity $\phi$ is expected because an isotropic thermal expansion would cause a proportional change in every linear dimension of the body (McTigue 1986).

Assuming incremental linear thermo-elasticity, the total variation of volume of the porous material is obtained as the sum of the volume change of each sub-problem.

The variation of the Lagrangian porosity $n$ (equation (1)), can be written as the difference between the volume change of the element and the volume change of the solid phase:

$$\Delta n = \frac{\Delta V - \Delta V_s}{V_0} = \left[ -c_s \Delta u - c_d (\Delta\sigma - \Delta u) + \alpha_s \Delta T \right]$$
$$- \left[ -(1-n_0)c_s \Delta u - c_s (\Delta\sigma - \Delta u) + (1-n_0)\alpha_s \Delta T \right] \qquad (6)$$

Equation (6) can be simplified to obtain the following expression for the elastic change of porosity in drained condition:

$$\Delta n = -(c_d - c_s)\Delta\sigma + n_0 c_n \Delta u + n_0 \alpha_n \Delta T \qquad (7)$$

where $c_n$ and $\alpha_n$ are respectively the compressibility and the volumetric thermal expansion coefficient of pore-volume defined by the following expressions:

$$c_n = \frac{1}{n_0}\left[ c_d - (1+n_0)c_s \right] \qquad (8)$$

$$\alpha_n = \alpha_s \qquad (9)$$

Replacing $\phi = \phi_0 + \Delta\phi$ in the equation (3) and considering only the first order terms the variation of the Eulerian porosity can be written in the following form:

$$\Delta\phi = \Delta n - \phi_0 \frac{\Delta V}{V_0} \qquad (10)$$

In equation (10) $\Delta n$ is given by equation (7) and $\Delta V/V_0$ is the sum of the variations of volume of the three sub-problems presented in the Figure (1). With $\phi_0 = n_0$ and using equations (8) and (9) the following expression is found for the variation of the Eulerian porosity, which is equivalent to the expression given by McTigue (1986) for an ideal porous material:

$$\Delta\phi = -\left[c_d\left(1-\phi_0\right)-c_s\right]\left(\Delta\sigma - \Delta u\right) \qquad (11)$$

This equation clearly highlights that, as mentioned above, a drained temperature change does not affect the Eulerian porosity in case of an ideal porous material.

For undrained thermo-mechanical loading (no change in the pore fluid mass), the volume change of the element is equal to the sum of the variations of volume of its constituents:

$$\begin{aligned}&-c_s\Delta u - c_d\left(\Delta\sigma - \Delta u\right) + \alpha_s\Delta T = \\ &-\left(1-n_0\right)c_s\Delta u - c_s\left(\Delta\sigma - \Delta u\right) + \left(1-n_0\right)\alpha_s\Delta T - n_0 c_f \Delta u + n_0 \alpha_f \Delta T\end{aligned} \qquad (12)$$

Equation (12) can be simplified to obtain the following expression for the pore pressure change due to a thermo-mechanical loading.

$$\Delta u = B\Delta\sigma + \Lambda\Delta T \qquad (13)$$

where $B$ is the Skempton's (1954) coefficient

$$B = \frac{\left(c_d - c_s\right)}{n_0\left(c_f - c_s\right) + \left(c_d - c_s\right)} \qquad (14)$$

and $\Lambda$ is the coefficient of thermal pressurization

$$\Lambda = \frac{n_0\left(\alpha_f - \alpha_s\right)}{n_0\left(c_f - c_s\right) + \left(c_d - c_s\right)} \qquad (15)$$

Using the equations (8) and (9), the thermal pressurization coefficient $\Lambda$ (equation (15)) can be expressed in terms of the thermal expansion coefficient and the compressibility of the pore-volume, $\alpha_n$ and $c_n$ respectively, and the expression given by Rice (2006) is retrieved:

$$\Lambda = \frac{\alpha_f - \alpha_n}{c_f + c_n} \qquad (16)$$

Equation (16) clearly highlights that the discrepancy between the thermal expansion of the pore fluid and that of the pore volume is the factor causing the thermal pressurization of porous materials.

Similarly, the volumetric strain of the element $\varepsilon_v$ (positive in contraction) is obtained as the sum of the variations of volume in the three sub-problems per unit total volume (in reference state):

$$\varepsilon_v = -\frac{\Delta V}{V_0} = c_s\Delta u + c_d\left(\Delta\sigma - \Delta u\right) - \alpha_s\Delta T \qquad (17)$$

By replacing $\Delta u$ with the expression given in equation (13), one finds the following expression for the volumetric deformation of a porous material subjected to a thermo-mechanical loading:

$$\varepsilon_v = c_u\Delta\sigma - \alpha_u\Delta T \qquad (18)$$

where $c_u$ and $\alpha_u$ are respectively the undrained compressibility and the undrained volumetric thermal expansion coefficient of porous material expressed by the following equations:

$$c_u = c_d - B\left(c_d - c_s\right) \qquad (19)$$

$$\alpha_u = \alpha_s + \Lambda\left(c_d - c_s\right) \qquad (20)$$

Using the equations (14) and (15) the expression of the undrained thermal expansion coefficient of a thermo-elastic porous material given by McTigue (1986) is retrieved.

$$\alpha_u = \alpha_s + Bn_0\left(\alpha_f - \alpha_s\right) \tag{21}$$

# 3 Thermal pressurization of granular rocks

The phenomenon of thermal pressurization has been studied experimentally on Rothbach sandstone. The Rothbach sandstone has a porosity of 16%. Its composition is 85% quartz, 12% feldspars and 3% clay.

## *3.1 Experimental setting*

A schematic of the triaxial cell used in this study is shown in Figure (2). This cell can sustain a confining pressure up to 60 MPa. It contains a system of hydraulic self-compensated piston. The loading piston is then equilibrated during the confining pressure build up and directly applies the deviatoric stress. The axial and radial strains are measured directly on the sample inside the cell with two axial transducers and four radial ones of LVDT type (see section AA on Figure (2)). These internal devices allow to avoid the main errors of strain measurements of devices external to the cell such as the compliance of the loading device, the tilting of the specimen and the bedding errors at the ends of the specimen. The confining pressure is applied by a servo controlled high pressure generator. Hydraulic oil is used as confining fluid. The pore pressure is applied by another servo-controlled pressure generator with possible control of pore volume or pore pressure.

The heating system consists of a heating belt around the cell which can apply a temperature change with a given rate and regulate the temperature, and a thermocouple which measures the temperature of the heater. In order to limit the temperature loss, an insulation layer is inserted between the heater element and the external wall of the cell. A second insulation element is also installed beneath the cell. The heating system heats the confining oil and the sample is heated consequently. Therefore there is a discrepancy between the temperature of the heating element in the exterior part of the cell and that of the sample. In order to control the temperature in the centre of the cell, a second thermocouple is placed at the vicinity of sample. The temperature given by this transducer is considered as the sample temperature in the analysis of the test results.

## *3.2 Effect of mechanical and thermal deformation of the drainage system*

The undrained condition is defined as a condition in which there is no change in the fluid mass of the porous material. Achieving this condition in a conventional triaxial system is very difficult. In these systems an undrained test is usually performed by closing the valves of the drainage system. As the drainage system has a non-zero volume, it experiences volume changes due to its compressibility and its thermal expansivity. A variation of the volume in the drainage system induces a fluid flow into or out of the sample to achieve pressure equilibrium between the sample and the drainage system. This fluid mass exchange between the sample and the drainage system and more generally the mechanical and thermal deformations of the drainage system modify the measured pore pressure during the test.

Wissa (1969) was the first who studied this problem for a mechanical undrained loading. He presented an expression for the measured pore pressure increase as a function of the compressibilities of water, soil skeleton, pore-water lines and pressure measurement system, but the compressibility of the solid phase was not considered in this work. Bishop (1976) completed the formulation of Bishop (1973) and presented an extension of the work of Wissa (1969) taking into account the compressibility of the solid grains. The proposed expression was first used by Mesri et al. (1976) to correct the measured pore pressure in undrained isotropic compression tests.

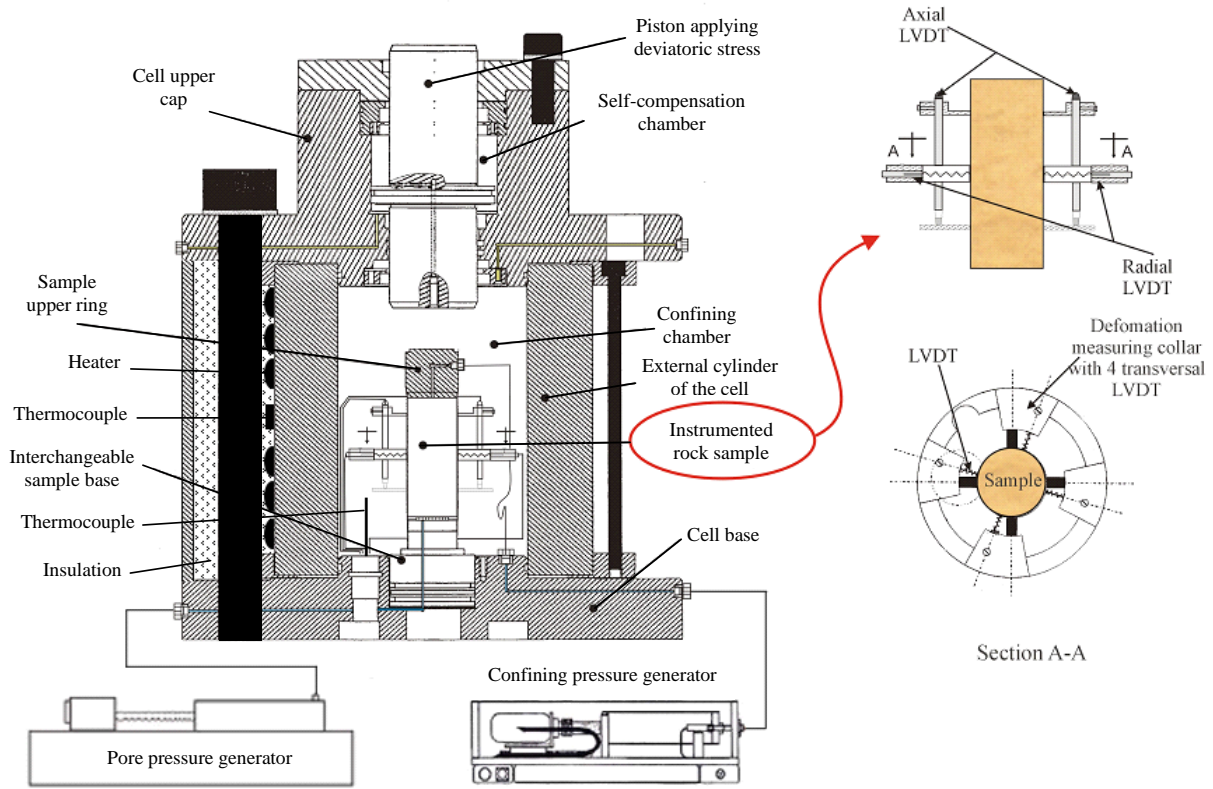

**Figure 2: Triaxial cell**

### 3.2.1 Analysis of the correction terms

We here present an extension of the work of Bishop (1976) to take into account the mechanical compressibility and the thermal expansivity of the drainage and pressure measurement system on the measured pore pressure increase during an undrained thermo-mechanical loading. The same approach as the one used in section (2) for the derivation of the governing equations can be used here by taking into account the drainage system. The problem is schematized in Figure (3) and the volume changes for the porous material element, its constituents and the drainage system are presented in table (2).

Four additional parameters are introduced in the problem to take into account the influence of the drainage system. $V_L$ is the volume of fluid in the drainage system. $c_L$ and $\alpha_L$ are the compressibility and the thermal expansion coefficient of the drainage system respectively defined as the variation of the volume of drainage system due to a unit variation in pore pressure and temperature. The parameter $c_L$ is equivalent to $C_L + C_M$ in Wissa (1969) and Bishop (1976) who give some typical values of this parameter which varies between 0.08 and 0.3 mm$^3$/MPa.

In most triaxial devices, the drainage system can be separated into two parts, one situated inside the triaxial cell and the other one situated outside the cell. In the part inside the cell, one can assume that the temperature change $\Delta T$ is the same as the one of the sample; in the other part situated outside the cell, the temperature change is smaller than $\Delta T$ and varies along the drainage lines. Let us define an equivalent temperature change $\Delta T_L$ such that the volume change of the entire drainage system caused by $\Delta T_L$ is equal to the volume change induced by the real non homogeneous temperature field. The temperature ratio $\beta = \Delta T_L / \Delta T$ is an additional parameter which is evaluated on a calibration test as explained further (see section 3.2.2). As the thermal expansion coefficient and the compressibility of water both vary with temperature, the parameters used for the water in the drainage system, $\alpha_{fL}$ and $c_{fL}$, are different from the parameters used for the pore fluid of the rock.

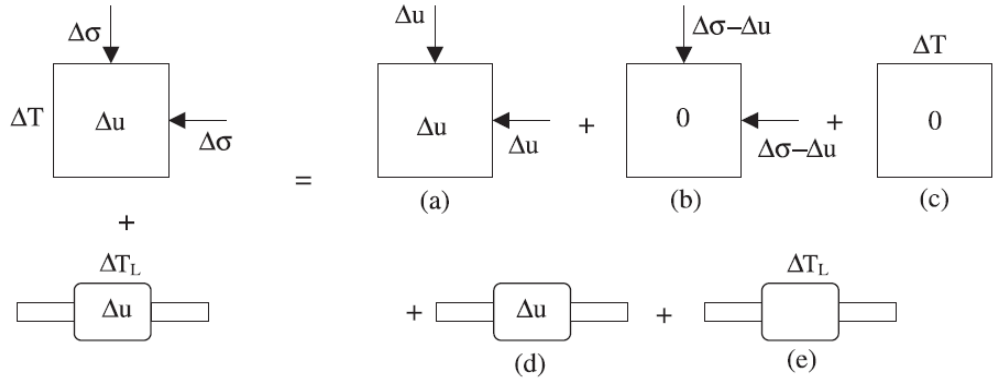

**Figure 3: Decomposition of the problem of a thermo-mechanical loading with taking into account the drainage and pressure measurement system**

| Problem | Variation of volume of constituents | | Variation of volume of element |
|---|---|---|---|
| | Pore fluid | Solid phase | |
| (a) | $-n_0 V_0 c_f \Delta u$ | $-(1-n_0)V_0 c_s \Delta u$ | $-V_0 c_s \Delta u$ |
| (b) | - | $-V_0 c_s (\Delta\sigma - \Delta u)$ | $-V_0 c_d (\Delta\sigma - \Delta u)$ |
| (c) | $n_0 V_0 \alpha_f \Delta T$ | $(1-n_0)V_0 \alpha_s \Delta T$ | $V_0 \alpha_s \Delta T$ |
| | Fluid in drainage system | Drainage system | |
| (d) | $-V_L c_{fL} \Delta u$ | $c_L \Delta u$ | - |
| (e) | $V_L \alpha_{fL} \beta \Delta T$ | $\alpha_L \beta \Delta T$ | - |

**Table 2: Variations of volume of the porous material, its constituents and the drainage and pressure measurement system for the independent problems presented in Figure (3)**

For undrained condition, as mentioned before, the overall volume change of an element is the sum of the volume changes of its constituents. The volume change of the pore fluid must be decreased by a correction term corresponding to volume change of the drainage system. This correction term can be evaluated as the change of volume of the drainage system minus the change of volume of water in the drainage system, and is expressed as $V_L c_{fL} \Delta u - V_L \alpha_{fL} \beta \Delta T + c_L \Delta u + \alpha_L \beta \Delta T$. Consequently, the following expression is obtained for the measured pore pressure due to a thermo-mechanical loading:

$$\Delta u = B_{mes} \Delta\sigma + \Lambda_{mes} \Delta T \tag{22}$$

$B_{mes}$ and $\Lambda_{mes}$ are the measured Skemptons's coefficient and thermal pressurization coefficient respectively.

$$B_{mes} = \frac{(c_d - c_s)}{n_0 (c_f - c_s) + (c_d - c_s) + \dfrac{V_L c_{fL} + c_L}{V_0}} \tag{23}$$

$$\Lambda_{mes} = \frac{n_0(\alpha_f - \alpha_s) + \dfrac{\beta(V_L \alpha_{fL} - \alpha_L)}{V_0}}{n_0(c_f - c_s) + (c_d - c_s) + \dfrac{V_L c_{fL} + c_L}{V_0}} \tag{24}$$

Using the equations (8) and (9) the measured thermal pressurization coefficient can be expressed in terms of the compressibility and the thermal expansion coefficient of the pore volume:

$$\Lambda_{mes} = \frac{\alpha_f - \alpha_n + \dfrac{\beta(V_L \alpha_{fL} - \alpha_L)}{n_0 V_0}}{c_f + c_n + \dfrac{V_L c_{fL} + c_L}{n_0 V_0}} \tag{25}$$

The (corrected) Skempton's and thermal pressurization coefficients (equations (14) and (16)) can thus be expressed from the measured ones as:

$$B_{cor} = \frac{1}{\dfrac{1}{B_{mes}} - \dfrac{V_L c_{fL} + c_L}{V_0(c_d - c_s)}} \tag{26}$$

$$\Lambda_{cor} = \frac{\Lambda_{mes}}{1 + \dfrac{\beta(V_L \alpha_{fL} - \alpha_L)}{n_0 V_0 (\alpha_f - \alpha_n)} - \Lambda_{mes} \dfrac{V_L c_{fL} + c_L}{n_0 V_0 (\alpha_f - \alpha_n)}} \tag{27}$$

Equation (26) is the same as the one given by Bishop (1976) for the correction of the measured pore pressure during a mechanical loading.

The correction method proposed here is quite simple. It is applied directly on the results of the test and only requires two simple calibration tests (see next section), but it is restricted to elastic response of the sample and of the drainage system. It differs from the method proposed by Lockner and Stanchits (2002) who have modified the procedure of the test itself by imposing a computer-generated virtual 'no-flow boundary condition' at the sample-endplug interface to insure that no volume change occurs in the drainage system.

### 3.2.2 Calibration of the correction parameters

The volume of fluid in the drainage system, $V_L$, can be measured directly or estimated by using the geometrical dimensions of the drainage system. For the triaxial cell used in the present study, $V_L = 2300\,mm^3$. The compressibility of the drainage and pressure measurement systems $c_L$ is evaluated by applying a fluid pressure and by measuring the corresponding volume change in the pore pressure generator. A metallic sample is installed inside the cell to prevent the fluid to go out from the drainage system. Fluid mass conservation is written in the following equation which is used to calculate the compressibility $c_L$ of the drainage system:

$$\Delta V_L = (c_L + V_L c_f) \Delta u \tag{28}$$

$\Delta u$ and $\Delta V_L$ are respectively the pore pressure and the volume change measured by the pressure generator. For a single measurement, the volume change $\Delta V_L$ accounts also for the compressibility of the pressure generator and of the lines used to connect the generator to the main drainage system. To exclude the compressibility of these parts, a second measurement is done only on the pressure generator and the connecting lines. The volume change $\Delta V_L$ used in equation (28) is the difference between these two measurements. The compressibility of the drainage system, $c_L$, is evaluated equal to 0.27 mm$^3$/MPa.

The parameter $\beta$ and the thermal expansivity of the drainage system $\alpha_L$ are evaluated using an undrained heating test which is performed using a metallic sample with the measurement of the fluid pressure change in the drainage system. For the metallic sample $n_0 = 0$ and $c_d = c_s$ so that equation (24) is reduced to the following expression:

$$\Lambda_{mes} = \frac{\beta(V_L \alpha_{fL} - \alpha_L)}{V_L c_{fL} + c_L} \tag{29}$$

The thermal expansion coefficient $\alpha_{fL}$ and the compressibility $c_{fL}$ of water are known as functions of temperature and pressure (Spang 2002). As these variations are highly non-linear, the parameters $\beta$ and $\alpha_L$ cannot be evaluated directly but are back analysed from the calibration test results: the undrained heating test of the metallic sample is simulated analytically using equation (29) with a step by step increase of the temperature. For each step the corresponding water thermal expansion and compressibility are used. The parameters $\beta$ and $\alpha_L$ are back-calculated by minimizing the error between the measurement and the simulation results using a least-square algorithm. The test result and the back analysis are presented in Figure (4). The parameter $\beta$ is found equal to 0.46 and the thermal expansivity of the drainage system $\alpha_L$ for this test is found equal to 0.31 mm$^3$/°C.

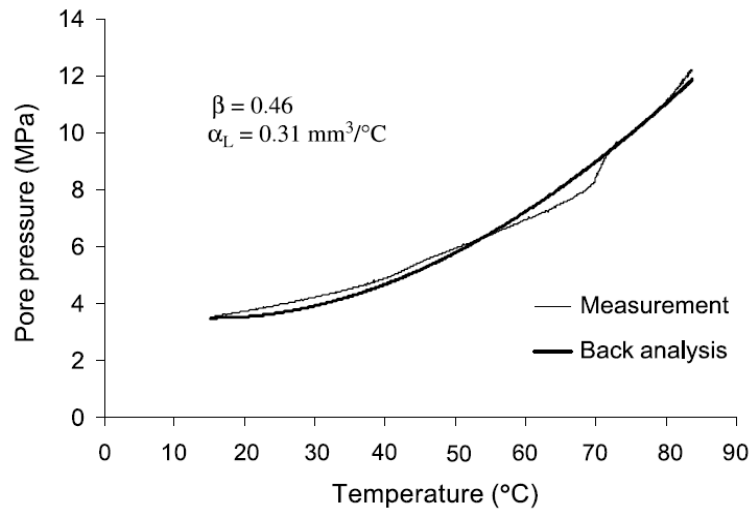

**Figure 4: Calibration test for evaluation of the temperature ration $\beta$ and the thermal expansion $\alpha_L$ of the drainage system-comparison with back analysis of the test**

### *3.3 Mechanical behaviour of Rothbach sandstone*

Equation (15) shows that to evaluate the thermal pressurization coefficient of a porous material, its drained compressibility and the compressibility of the solid phase, $c_d$ and $c_s$ respectively, must be known. In order to measure the compressibility of the solid phase, an isotropic compression test, called 'unjacketed test', was carried out in which equal increments of the confining pressure and the pore pressure were applied simultaneously, (see sub-problem (a) in Figure (1)). The compression modulus measured in this test, called the unjacketed modulus, is equal to the compression modulus of the solid grains for an ideal porous material. In the case of in inhomogeneous porous material (i.e., composed of two or more solids) the unjacketed modulus is some average of the bulk modulus of the constituents (Berryman 1992). As explained by Berryman, what this average should be is generally not known, but can be estimated in some simple cases using homogenisation techniques. The stress-strain response of the performed test is presented in Figure (5). The unjacketed modulus of the Rothbach sandstone, $K_s = 1/c_s$, is obtained as the slope of this curve equal to 41.6 GPa. This value is identified to the compression modulus $K_s$ of the solid phase which appears in the above poromechanical formulation (see Table 1).

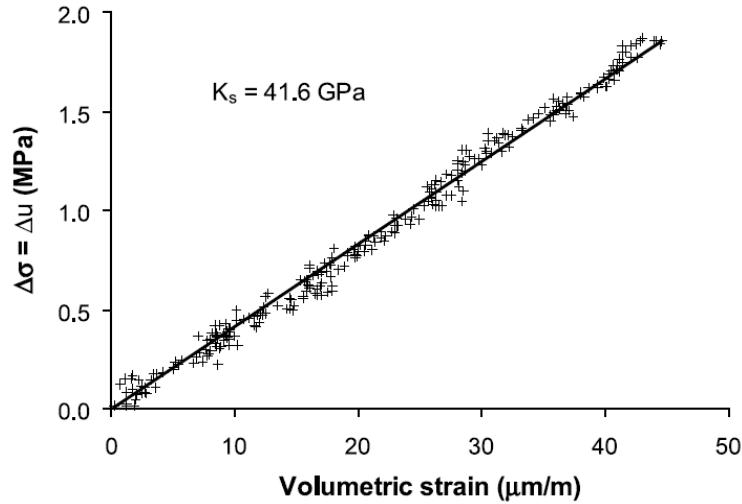

**Figure 5: Unjacketed test on Rothbach sandstone: evaluation of the compressibility of solid phase**

The unjacketed compressibility can also be estimated by knowing the mineralogy of the rock and the compressibility of each constituent. As mentioned above, the Rothbach sandstone contains 85% quartz, 12% feldspars and 3% clay. The values of elastic properties of different minerals can be found in Bass (1995). The compression modulus of quartz and the average compression modulus of feldspars are equal to 37.8 and 69.1 GPa respectively. The compression modulus of the clay solid grains is equal to 50 GPa (Skempton 1960, McTigue 1986). The homogenized unjacketed modulus can be evaluated using Hill's (1952) average formula:

$$K_s^{\text{hom}} = \frac{1}{2}\left[\sum f_i K_s^{(i)} + \left(\sum \frac{f_i}{K_s^{(i)}}\right)^{-1}\right] \tag{30}$$

where $f_i$ and $K_s^{(i)}$ are the volume fraction and the compression modulus of the i$^{\text{th}}$ constituent respectively. Using equation (30) the homogenized unjacketed modulus of Rothbach sandstone is evaluated equal to 41.1 GPa which confirms the measured value (41.6 GPa).

To measure the drained compressibility of the rock, an isotropic drained compression test with a loading and unloading cycle has been carried out. In this test, a fluid back pressure of 1MPa was maintained constant inside the sample. In the Figure (6) the volumetric strain response is shown versus the applied total stress. The observed non-linear response reflects the stress-dependent character of the rock compressibility (Zimmerman 1991, Sulem et al. 1999, 2006). Moreover this curve exhibits a different response in loading and unloading and irreversible deformation. Therefore the elastic tangent drained compression modulus, $K_d = 1/c_d$ is evaluated on the unloading curve.

For a porous material with a constant unjacketed modulus $K_s$ it can be shown theoretically that the stress-dependent drained bulk modulus $K_d$ depends on Terzaghi effective stress $\sigma'_{Terzaghi} = \sigma - u$.

Various theoretical demonstrations of this statement can be found in Zimmerman (1991), Berryman (1992) Coussy (2004) and Gurevich (2004) and from micromechanical considerations in Dormieux et al. (2002). Experimental verification is presented by Boutéca et al. (1994) for two sandstones. In the following, the results are analysed and presented in terms of Terzaghi effective stress.

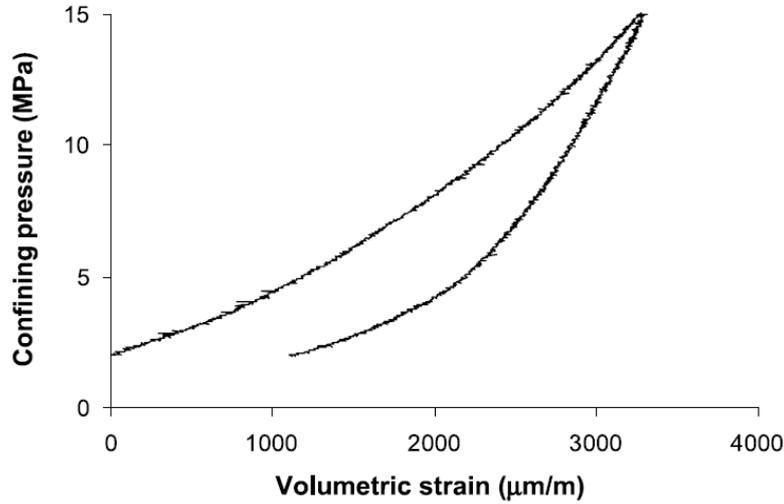

**Figure 6: Drained hydrostatic compression test on Rothbach sandstone**

The slope of the unloading curve is directly evaluated on the experimental data presented on Figure (6). For this test, the data were recorded every 2 seconds which corresponds approximately to 15 kPa of stress increment. In order to minimise the effect of small fluctuations of the recorded data on the resulting derivative, one can either fit the data with a mathematical function and calculate the derivative or evaluate the slope by a difference quotient calculated on a discrete set of $N$ points centred on the corresponding point. The choice of $N$ depends on the quality of the experimental data and on the amplitude of the fluctuations. The first method is commonly used and gives generally satisfactory results. However the mathematical form of the assumed fitting function obviously determines the derivative. Different mathematical functions can give equally good results for the fitting of given data, and result in different evaluations of the derivative. The second method based on discrete derivative calculus is used here. The slope $S$ of a $x-y$ curve is given by:

$$S = \frac{\sum x^{(i)} \sum y^{(i)} - N \sum x^{(i)} y^{(i)}}{\left(\sum x^{(i)}\right)^2 - N \sum \left(x^{(i)}\right)^2} \qquad (31)$$

where $x^{(i)}$ and $y^{(i)}$ are the coordinates of the i$^{\text{th}}$ point. To calculate the drained compression modulus $K_d$ the $x^{(i)}$ and $y^{(i)}$ are respectively $\varepsilon_v^{(i)}$ and $\sigma'^{(i)}$, the volumetric strain and effective stress data at the i$^{\text{th}}$ point. With $N = 61$ the resulted compression modulus plotted as a function of the effective stress curve is presented in Figure (7). Based on these results, the tangent compression modulus $K_d$ can be approximated with a bi-linear equation.

$$\begin{array}{ll} K_d = 0.96\sigma' + 0.70 & \sigma' \leq 9\,MPa \\ K_d = 0.07\sigma' + 8.72 & \sigma' > 9\,MPa \end{array} \quad \text{with } K_d \text{ in GPa and } \sigma' \text{ in MPa} \qquad (32)$$

Equation (32) expresses that the compression modulus increases significantly at relatively low effective stress (up to 9 MPa) and does not evolve much above this value. However, at higher stresses it is expected that pore collapse and grain crushing phenomena will induce a reduction of the compression modulus (Fortin et al. 2005).

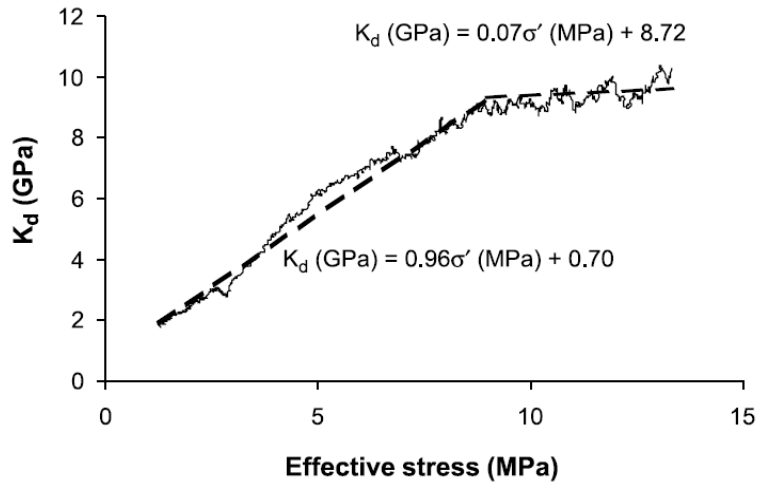

**Figure 7: Drained compression modulus of Rothbach sandstone as a function of the effective stress**

## *3.4 Thermal behaviour of Rothbach sandstone*

In order to measure the thermal expansion coefficient of the rock, a drained heating test (see sub-problem (c) in Figure (1)) was carried out under a constant confining pressure. In the case of an inhomogeneous porous material (i.e., composed of two or more solids), the variation of volume of the sample in this test is characterized by the drained thermal expansion coefficient $\alpha_d$. This coefficient can be expressed in terms of average over the porous medium volume (Palciauskas and Domenico 1982).

$$\alpha_d = (1 - n_0)\alpha_s + n_0 \alpha_n \tag{33}$$

The difference between the expansion coefficients $\alpha_d$ and $\alpha_s$ reflects the difference between the thermal response of the bulk porous medium and that of the solid phase alone (McTigue 1986). In practice, the coefficient $\alpha_d$ also takes into account the non-reversible thermal deformations which can be produced by the microfracture generation due to the discrepancy between the thermal expansions of different minerals within the rock (Palciauskas and Domenico 1982). According to the experimental study of Walsh (1973), microfracture generation is only initiated at elevated temperatures. The test results also show that the reversible component of the drained expansivity $\alpha_d$ is approximately equal to that of the solid phase $\alpha_s$. However, in practice, the expansion coefficient $\alpha_d$ is commonly used in the literature as a thermo-elastic coefficient (equivalent to $\alpha_b$ in Palciauskas and Domenico (1982) and $\alpha'_s$ in McTigue (1986)).

Using equation (9) in the equation (33) in the case of an ideal porous material, one finds $\alpha_d = \alpha_s$.

The initial temperature of the performed drained heating test was 21°C. The confining pressure and the pore water pressure during the test were maintained constant at 2.5 MPa and 1.0 MPa respectively. The drained thermal expansion coefficient is the slope of the temperature-volumetric strain response and is evaluated as $28 \times 10^{-6} \left(°C\right)^{-1}$ (Figure 8).

The coefficient of thermal expansion of the solid phase $\alpha_s$ can also be evaluated using homogenisation theory, knowing the mineralogy of the rock and the thermal expansion coefficients of its constituents. The following equation can be used to calculate the average thermal expansion coefficient of a two phase linear thermo-elastic composite material (Berryman 1995, Zaoui 2000):

$$\alpha_s^{\text{hom}} = \left(f_1 \alpha_s^{(1)} + f_2 \alpha_s^{(2)}\right) + \frac{\dfrac{1}{K_s^{\text{hom}}} - \left(\dfrac{f_1}{K_s^{(1)}} + \dfrac{f_2}{K_s^{(2)}}\right)}{\dfrac{1}{K_s^{(2)}} - \dfrac{1}{K_s^{(1)}}} \left(\alpha_s^{(2)} - \alpha_s^{(1)}\right) \quad (34)$$

where $\alpha_s^{(i)}$ is the thermal expansion coefficient of the i$^{\text{th}}$ constituent. The other parameters are introduced before in equation (30). Note that equation (34) reflects the thermo-mechanical coupling as the compressibility of the constituents and of the composite material affects its thermal expansion.

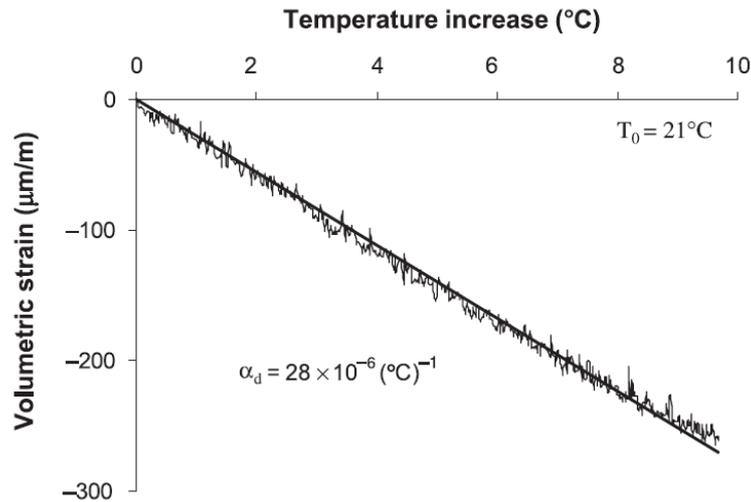

**Figure 8: Drained heating test on Rothbach sandstone**

The thermal expansion coefficient of quartz is equal to $33.4 \times 10^{-6} \left(°C\right)^{-1}$ (Palciauskas and Domenico 1982). The average thermal expansion coefficient of feldspars can be taken equal to $11.1 \times 10^{-6} \left(°C\right)^{-1}$ (Fei 1995). For the clay solid grains this coefficient is equal to $34 \times 10^{-6} \left(°C\right)^{-1}$ (McTigue 1986). Considering the small volume fraction of clay in Rothbach sandstone, the equation (34) can be used to calculate the homogenized thermal expansion coefficient of solid grains by neglecting this part and taking into account only quartz and feldspar minerals. Using the given parameters and the $K_s^{\text{hom}}$ calculated with equation (30), the thermal expansion coefficient of Rothbach sandstone is found equal to $29.7 \times 10^{-6} \left(°C\right)^{-1}$. This value is in very good accordance with the measured thermal expansion coefficient equal to $28 \times 10^{-6} \left(°C\right)^{-1}$.

The phenomenon of thermal pressurization was studied in an undrained heating test under constant isotropic stress equal to 10 MPa. The initial temperature of the sample was 20°C. The results are shown on Figures (9) and (10) where the measured pore pressure increase and the volumetric strain are plotted versus the temperature increase. The non-linearity of the observed thermal pressurization is due to the (effective) stress-dependent compressibility of the sandstone and also to the temperature and pressure dependent compressibility and thermal expansion of pore water.

For a pore pressure close to the confining pressure, the pressurization curve becomes almost horizontal and no more fluid pressurization is observed. This phenomenon is probably due to the presence of water between the sample and the rubber membrane when the difference between the confining pressure and the pore pressure is too small. At the beginning of the test the volumetric strain could not be recorded due to a failure of the displacement sensors inside the triaxial cell. Therefore the volumetric strain-temperature curve only starts from 40°C.

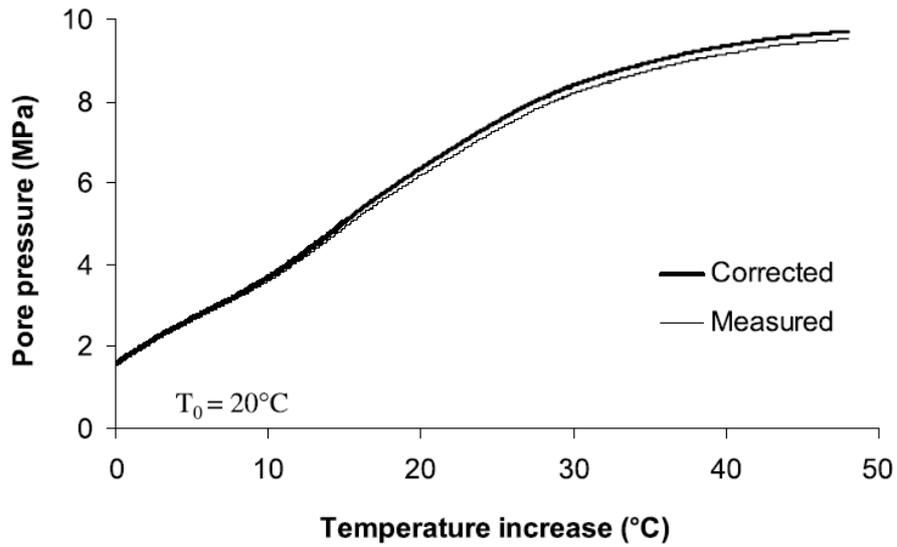

**Figure 9: Undrained heating test on Rothbach sandstone: pore pressure change versus temperature increase – measured and corrected values**

The coefficient of thermal pressurization is calculated for the measured temperature and pore pressure increase. Here again because of the small fluctuations of the experimental data, the derivative was calculated using equation (31). To calculate the thermal pressurization coefficient $\Lambda$ the $x^{(i)}$ and $y^{(i)}$ are respectively $T^{(i)}$ and $p^{(i)}$, the temperature and pore pressure increase data at the i$^{th}$ point. The calculated thermal pressurization for $N = 61$ is presented on the Figure (11) as a function of the temperature. We can see that the thermal pressurization coefficient $\Lambda$ increases up to about 0.25 MPa/°C and then decreases to 0.025 MPa/°C at the end of the test.

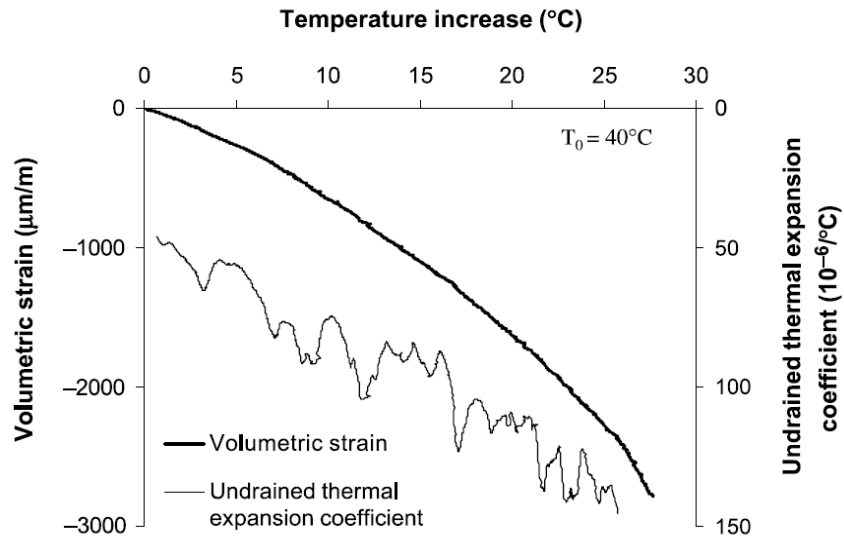

**Figure 10: Undrained heating test on Rothbach sandstone: volumetric strain and undrained thermal expansion coefficient versus temperature increase**

The undrained thermal expansion coefficient $\alpha_u$ is calculated using equation (31) replacing $x^{(i)}$ and $y^{(i)}$ respectively by $T^{(i)}$ and $\varepsilon_v^{(i)}$, the temperature and the volumetric strain data at the i$^{th}$ point. The result is presented in the Figure (10) and shows the increase of the undrained thermal expansion coefficient with the temperature. This important increase is mainly attributed to the increase of the thermal expansion coefficient of pore water with the temperature, shown in Figure (12).

## 3.5 Correction of measured pressurization

The observed pore pressure increase during the undrained heating test is corrected for the effect of drainage and pressure measurement system using equation (27). Because of the small fluctuations of the recorded data, this correction cannot be done directly on the coefficients calculated using equation (31). The measured temperature and pore pressure are first fitted with sixth degree polynomials as a function of data acquisition time and then the derivative of the pore pressure with respect to temperature (equal to the thermal pressurization coefficient $\Lambda$), is calculated (Figure 11). The comparison with the curve calculated using the equation (31) shows a good compatibility between the results. The corrected pressurization curve is then calculated and presented in the Figure (9). The corrected pore pressure is slightly greater than the measured one. However, due to the small volume of the drainage system, this difference is very limited. The corrected values of the thermal pressurization coefficient are also presented in Figure (11).

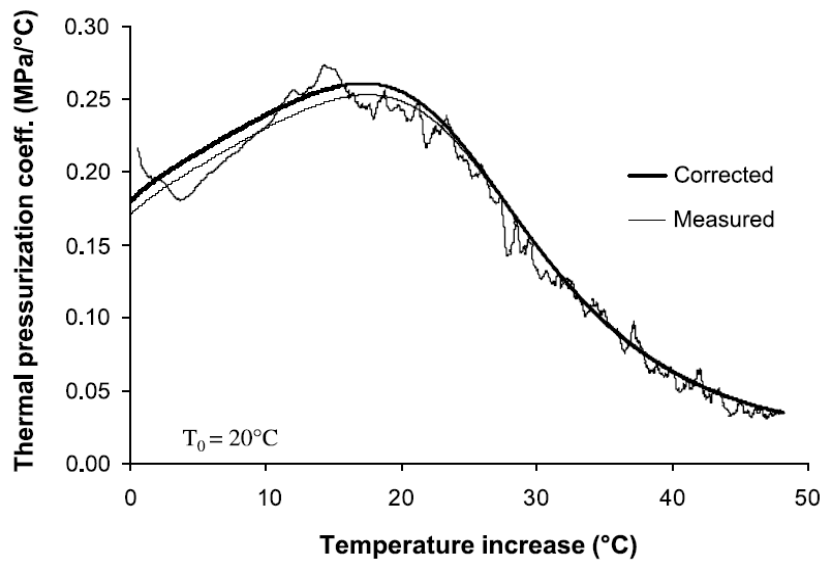

**Figure 11: Undrained heating test on Rothbach sandstone: thermal pressurization coefficient versus temperature increase – measured and corrected values**

# 4  Analytical simulation of the undrained heating test

The above analysis is validated through the simulation of the undrained heating test and comparison with the experimental results.

The formulation presented for the undrained thermo-elastic behaviour of porous materials along with the non-linear elastic constitutive model developed for the Rothbach sandstone enable us to simulate analytically the undrained heating test. As the test was performed under a constant confining pressure, the term $\Delta\sigma$ in the equations (13) and (18) is null. The pore pressure increase, while the confining pressure is maintained constant, induces a decrease of the effective stress (elastic unloading) and thus the expression of the compression modulus in unloading (equation (32)), must be used to calculate $c_n$ in equation (16). It is assumed that the rock compressibility is not affected by temperature changes. The volumetric strains were calculated using equation (18). The thermal expansion coefficient and the compressibility of water depend upon the pore pressure and the temperature and this dependency is taken into account (Spang 2002). Figures (12) and (13) present respectively the variations of thermal expansion coefficient and compressibility of water with temperature for different pressures.

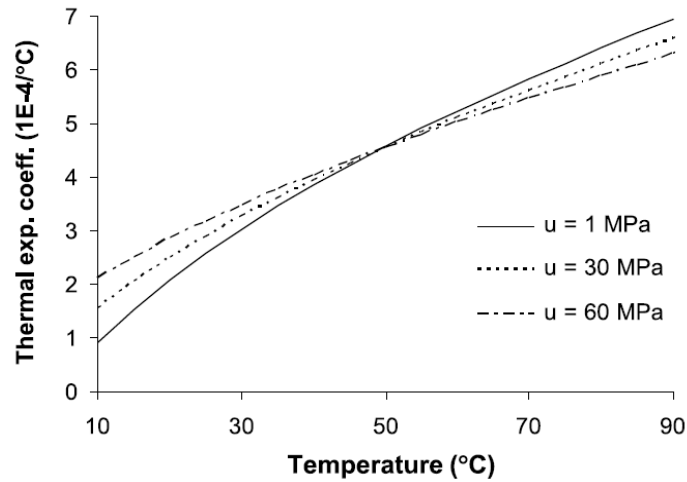

Figure 12: **Variation of the thermal expansion coefficient of water with temperature and pressure**

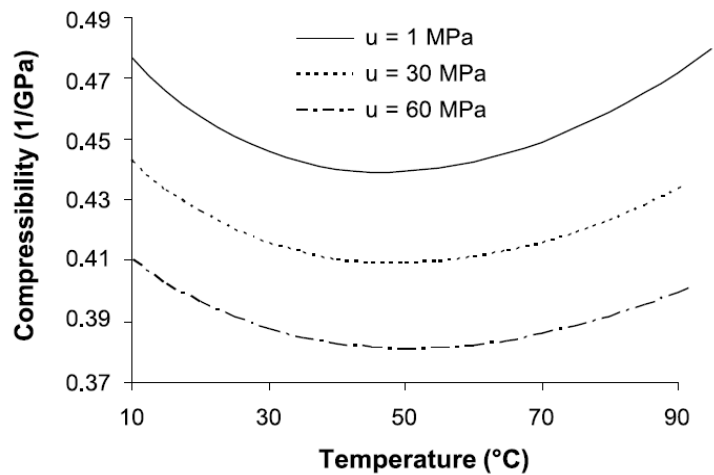

Figure 13: **Variation of the compressibility of water with temperature and pressure**

The results of simulation presented on Figures (14), (15), (16) and (17) show a good agreement with the results of undrained heating test. The phenomenon of thermal pressurization and also the undrained thermal expansion of the rock are well reproduced using the results of the drained isotropic compression and drained heating tests. This analytical simulation of the undrained heating test clearly shows the non-linear and (effective) stress-dependent character of thermal pressurization. Among the four parameters of equation (16), $\alpha_n$ is equal to $28 \times 10^{-6} \left(°C\right)^{-1}$ and remains constant during the simulation. $c_f$ varies slightly between 0.45 GPa$^{-1}$ and 0.43 GPa$^{-1}$. Therefore the variation of the thermal pressurization coefficient $\Lambda$ is mainly controlled by the variations of $\alpha_f$ and $c_n$. The thermal expansion of water $\alpha_f$ varies with temperature and pressure and increases from $2.09 \times 10^{-4} \left(°C\right)^{-1}$ to $5.64 \times 10^{-4} \left(°C\right)^{-1}$ during the simulation. The compressibility of pore-volume $c_n$ is calculated using equation (8). The compressibility of solid phase $c_s$ is considered constant and the drained compressibility of rock $c_d$ increases when the effective stress decreases (equation (32)). During the test, the pore pressure increase while the confining pressure is maintained constant, induces a decrease of the effective stress. Consequently the compressibility of pore-volume $c_n$ increases during the simulation from 0.47 GPa$^{-1}$ in the beginning of the simulation to 8.52 GPa$^{-1}$ at the end for a zero effective stress. In the beginning of the test the increase of $\alpha_f$ with the temperature dominates and the thermal pressurization coefficient $\Lambda$ increases up to about 0.24 MPa/°C. Then the important increase of $c_n$ decreases $\Lambda$ to 0.06 MPa/°C at the end of the test.

As mentioned above, during the undrained heating test, the observed pore pressure cannot reach the confining pressure. However analytical simulation makes it possible to predict the critical temperature for which the pore pressure reaches the confining pressure: 68°C for this test.

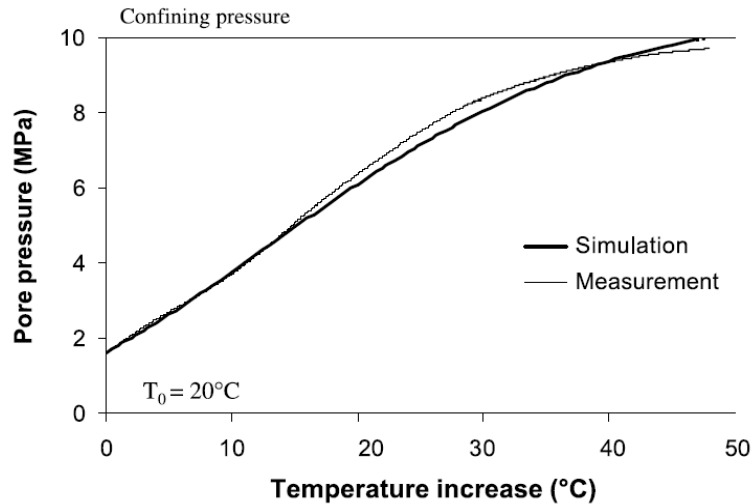

**Figure 14: Analytical simulation of the undrained heating test on Rothbach sandstone, comparison with the measured pore pressure**

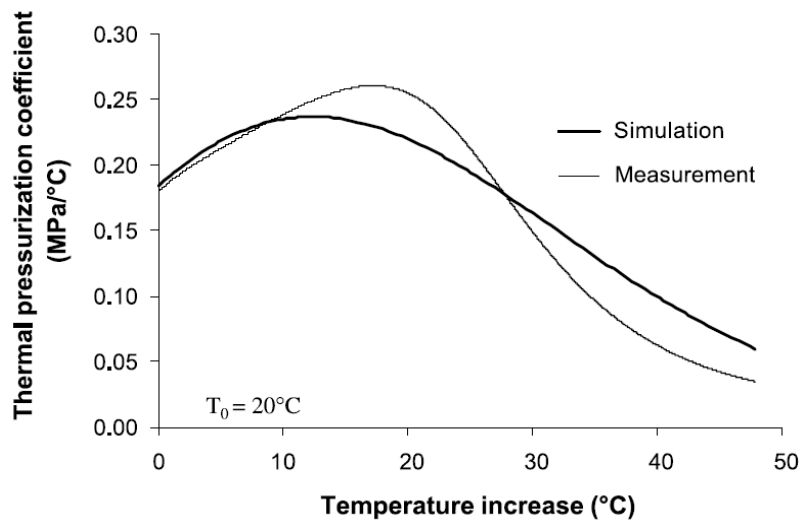

**Figure 15: Analytical simulation of the results of the undrained heating test on Rothbach sandstone, comparison with the measured thermal expansion coefficient**

# 5  Stress and temperature dependency of the thermal pressurization coefficient

In the previous sections the temperature and stress-dependent character of the thermal pressurization coefficient is clearly shown by the experimental results of the undrained heating test and the analytical simulation of this test. On one hand the temperature dependency of this coefficient is mainly due to the increase of the thermal expansion coefficient and in a lesser extent to the change of water compressibility with temperature. In some geomaterials the parameters $c_d$ and $c_s$ and consequently the pore-volume compressibility $c_n$ (equation (8)) are also temperature sensitive which will also influence the coefficient $\Lambda$. On the other hand the stress-dependency of the thermal pressurization coefficient is also due to the stress-dependent character of the drained compressibility $c_d$ and consequently of the pore-volume compressibility $c_n$ of the porous material, as presented above for the Rothbach sandstone.

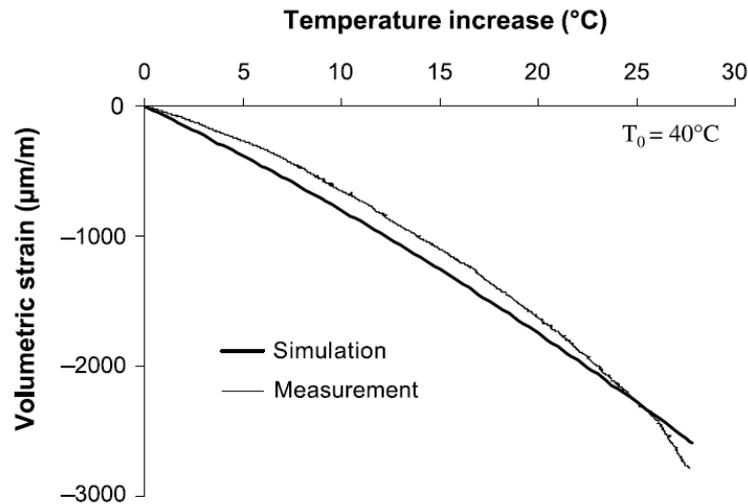

**Figure 16: Analytical simulation of the results of the undrained heating test on Rothbach sandstone, comparison with the measured volumetric strain**

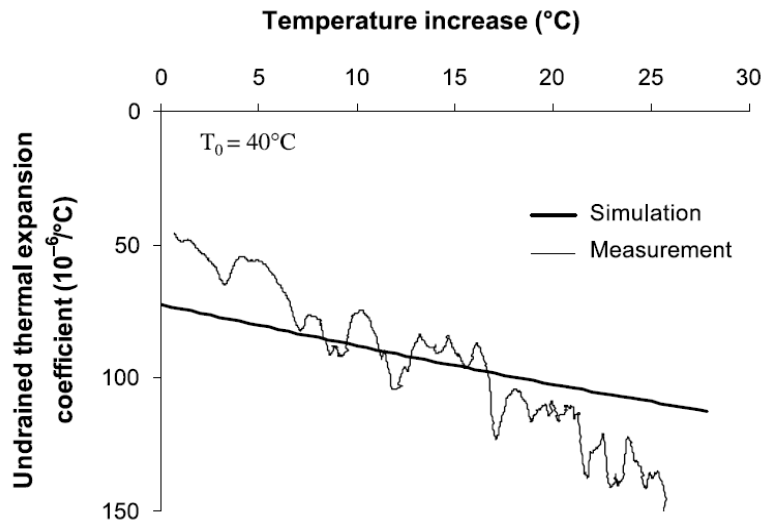

**Figure 17: Analytical simulation of the results of the undrained heating test on Rothbach sandstone, comparison with the measured undrained thermal expansion coefficient**

The first measurement of the thermal pressurization coefficient for sandstone is probably due to Campanella and Mitchell (1968) who also observed the stress-dependency of this coefficient. Applying a temperature increase from 5°C to 15°C, the measured $\Lambda$ coefficient for 0.25 MPa and 0.58 MPa effective stresses was found respectively equal to 0.02 MPa/°C and 0.05 MPa/°C.

It is common in the literature to use a single average value for the thermal pressurization coefficient. This average value is calculated using the average values of $\alpha_f$ and $c_f$ for a considered temperature range, and constant $\alpha_n$ and $c_n$ coefficients. For example, Palciauskas and Domenico (1982) used the data provided by Dropek et al (1978) to estimate the coefficient of thermal pressurization of Kayenta sandstone. With $n = 0.2$, $K_d = 9.52 GPa$, $K_s = 39.1 GPa$ and the average water properties between 10°C to 90°C, the coefficient $\Lambda$ was estimated equal to 0.59 MPa/°C.

Using the non-linear elastic model presented in equation (32), the measured values of $\alpha_s$ and $c_s$ and knowing the compressibility and thermal expansion coefficient of water as functions of temperature and pore pressure, we can evaluate the thermal pressurization coefficient $\Lambda$ of Rothbach sandstone as a function of temperature and effective stress (equation (16)) and obtain a good accordance with the undrained heating test as shown above. The calculated coefficient $\Lambda$ is presented on Figure (18) as a function of the effective stress up to 15 MPa and for different temperatures from 20°C to 90°C. The results show clearly the (effective) stress and temperature dependent character of the thermal pressurization coefficient. The stress-dependency is significant in the range of stress-dependency of the rock compressibility (up to 9 MPa). The values of the coefficient $\Lambda$ vary from 0.02 to 0.72 MPa/°C depending on the effective stress and the temperature.

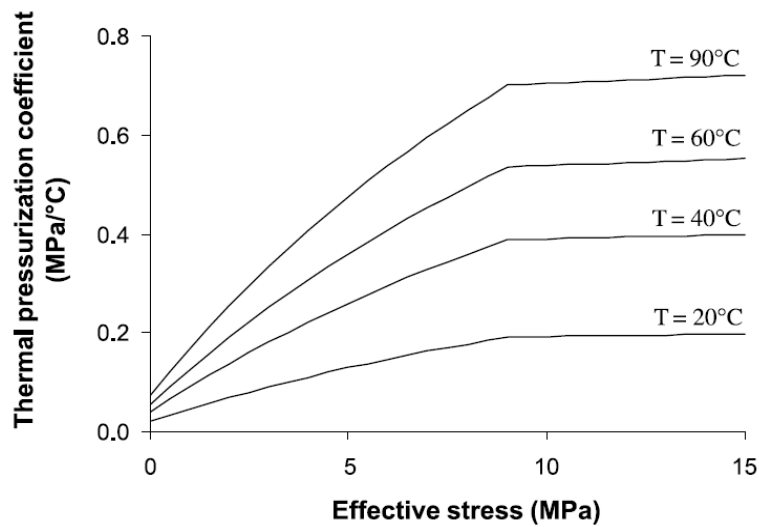

**Figure 18: Variation of thermal pressurization coefficient of Rothbach sandstone with effective stress and temperature**

# 6 Conclusion

The phenomenon of thermal pressurization is studied theoretically and experimentally for a saturated granular rock, the Rothbach sandstone. It has been demonstrated that this phenomenon is controlled, on one hand by the discrepancy between the thermal behaviour of the pore fluid and of the solid phase, and on the other hand by the compressibility of the pore volume. The strong influence of stress and temperature on thermal pressurization of rocks is explained by the stress-dependent character of the compressibility of porous rocks and by the effect of temperature and pressure on the thermal and mechanical properties of water.

The experimental study has consisted first in the characterisation of the mechanical and thermal response of the solid phase of the rock. A stress-dependent elastic model has been calibrated on a drained compression test. The experimental results of an undrained heating test showed clearly the stress and temperature dependent character of the thermal pressurization coefficient. In the analysis of the tests data, the effect of mechanical and thermal deformation of the drainage system has been accounted for. The results of the undrained heating test have been theoretically simulated and a good agreement with the experimental results has been obtained. This approach gives confidence in the evaluation of the thermal pressurization coefficient for various conditions of temperature and stress level.

# 7 Acknowledgments

The authors wish to thank Sylvine Guédon and Ahmad Pouya for fruitful discussions. They wish also to thank François Martineau for his assistance in the experimental work.